\title{Efficient Construction of Broadcast Graphs}
\author{
A. Averbuch
\thanks{
School of Computer Sciences, Tel
Aviv University, Tel Aviv 69978, Israel }
\and R. Hollander Shabtai
\thanks{
School of Computer Sciences, Tel
Aviv University, Tel Aviv 69978, Israel and Afeka College  of Engineering, Tel-Aviv 69460, Israel }
\and Y. Roditty
\thanks{
School of Computer Sciences, Tel Aviv University, Tel Aviv 69978,
Israel and School of Computer Sciences, The Academic College of
Tel-Aviv-Yaffo, Tel-Aviv 61161, Israel. email: jr@post.tau.ac.il} }

\date{} 

\documentclass[a4paper,11pt]{article}

\usepackage{latexsym}
\usepackage{graphicx}
\usepackage{amsmath}
\usepackage{amssymb}
\usepackage{amsthm}
\usepackage{float}
\usepackage{amsmath,amssymb,graphics,graphicx,float,subfigure}

\newtheorem{lemma}{Lemma}[section]

\newtheorem{theorem}{Theorem}[section]

\newtheorem{definition}{Definition}[section]
\newcommand{\ignore}[1]{}

\begin{document}
\maketitle
\begin{abstract}
A broadcast graph is a connected graph, $G=(V,E)$, $ |V |=n$, in which each vertex can complete broadcasting of one message within at most $t=\lceil \log n\rceil$ time units. A minimum broadcast graph on $n$ vertices is a broadcast graph with the minimum number of edges over all broadcast graphs on $n$ vertices. The cardinality of the edge set of such a graph is denoted by $B(n)$.
In this paper we construct a new broadcast graph with
         $B(n) \le (k+1)N -(t-\frac{k}{2}+2)2^{k}+t-k+2$, for $n=N=(2^{k}-1)2^{t+1-k}$ and
         $B(n) \le (k+1-p)n -(t-\frac{k}{2}+p+2)2^{k}+t-k -(p-2)2^{p}$, for $2^{t} < n<(2^{k}-1)2^{t+1-k}$, where $t \geq 7$, $2 \le k \le \lfloor t/2 \rfloor -1$ for even $n$ and $2 \le k \le \lceil t/2 \rceil -1$ for odd $n$, $d=N-n$, $x= \lfloor \frac{d}{2^{t+1-k}} \rfloor$ and $ p = \lfloor \log_{2}{(x+1)} \rfloor$ if $x>0$ and $p=0$ if $x=0$.

  The new bound is an improvement upon the bounds appeared in \cite{bf},\cite{gf} and \cite{hl} and the recent bound presented by Harutyunyan and  Liestman (\cite{hln}) for odd values of $n$.


\end{abstract}

\section{Introduction}
\emph{Broadcasting} is an information distribution problem in a connected graph, in which one vertex, called the \emph{originator}, has to distribute a message to all other vertices by placing a series of calls among the communication lines of the graph. Once informed, the informed vertices aid the originator in distributing the message. This is assumed to take place in discrete  time units. The broadcasting has to be completed within a minimal number of time units subjected to the following constraints:


1. Each call involves only one informed vertex and one of its uninformed

           neighbors.

2. Each call requires one time unit.

3. A vertex can participate in at most one call at each time unit.

4. At each time unit many calls can be performed in parallel.

Formally, any network can be modeled as a simple connected graph $G=(V,E)$, $|V|=n$, where $V$ is the set of vertices and $E$ is the set of edges (the communication  lines).
For a given originator vertex, $u$, the \emph{broadcast time} of $u$, $b(u)$, is defined
as the minimum number of time units needed to complete broadcasting from $u$. Note that for any vertex $u \in V$,
$b(u)\geq \lceil \log n \rceil$ (to the sequel the base of logs is always 2), since at each time unit the number of informed vertices can at most double. The broadcast time $b(G)$ of the graph $G$ is defined as $max\{b(u)|u \epsilon G\}$ and  $G$ is called a broadcast graph if $b(G)= \lceil \log n \rceil$.

The \emph{broadcast number $B(n)$} is the minimum number of edges in any broadcast graph on $n$ vertices. A \emph{minimum broadcast graph (mbg)} is a broadcast graph on $n$ vertices with $B(n)$ edges.
Currently, the exact values of $B(n)$ are known only for $n=2^{p}$, $n=2^{p}-2$, $n=127$, and for several values of $n\leq63$, as detailed below.
Farley et al. \cite{fh} determined the values of $B(n)$ for $ n \leq 15$ and showed that
hypercubes are \emph{mbgs} such that $B(2^{p})=p2^{p-1}$ for any $p \ge 2$.
Mitchell and Hedetniemi \cite{mh} determined the value of $B(17)$,
while Bermond, Hell, Liestman and Peters \cite{bh} determined the values of
$B(n)$ for $ n = 18,19,30,31$.
Khachatrian and Haroutunian \cite{lh} and
independently Dinnen, Fellows and Faber \cite{dn} proved that
$B(2^{p}-2)=(p-1)(2^{p-1}-1)$ for all $p \geq 2$.

Since \emph{mbg's} seem to be difficult to find, many authors have devised methods to construct broadcast graphs. The number of edges in any broadcast graph on $n$ vertices gives an upper bound on $B(n)$. Several papers have shown methods to construct broadcast graphs by forming the compound of two known broadcast graphs (see \cite{bf}, \cite {dv}, \cite{hl} and \cite{lh}). These methods have proven effective for graphs on $n_{1} n_{2}$ vertices from two known broadcast graphs on $n_{1}$ and $n_{2}$ vertices. Thus, compounding produces good upper bound on $B(n)$ for many values of $n$. In particular, a very tight upper bound was obtained for $n=2^{p} - 2^{k}$ by compounding mbg's on $2^{k-1}$ and $2^{p-k+1}-2$ vertices: $B(2^{p}-2^{k}) \le \frac{2^{p}-2^{k}}{2}(p-\frac{k+1}{2})$ (see \cite{bf},\cite{lh}).

Broadcast graphs on other sizes can sometimes be formed by adding or deleting vertices from known broadcast graphs(see \cite{bh} for example). An efficient vertex addition method is suggested in \cite{hh}. The authors in \cite{hl} presented a method based on compounding and then merging several vertices into one that allows the construction of the best broadcast graphs  for almost all values of $n$, including many prime numbers.
In particular, a very tight upper bound on $B(n)$ is
$B(2^{p}-2^{k}+1)\leq 2^{p-1}(p-\frac{k}{2})$ (again by compounding mbg's on $2^{k}$ and $2^{p-k}$ vertices and then merging $2^{k}$ vertices into one).

Farley (\cite{af}) proposed the recursive method to construct minimal broadcast graphs and proved the general upper bound
$$B(n) \leq \frac{n\lceil \log n \rceil}{2} , ~2^{p-1}<n\leq 2^{p}.
\eqno(1)
$$
Other general upper bounds on $B(n)$ are obtained from a direct construction using binomial trees
(see \cite{gp},\cite{hl},\cite{lh}) for some values of $n$.

Direct construction of broadcast graphs is a difficult problem. The best upper bound from a direct construction for any $n$ is
$$B(n)\leq n(p-k+1)-2^{p-k}-\frac{1}{2}(p-k)(3p+k-3)+2k, \eqno(2)$$ where $n=2^{p}-2^{k}-r$, $0 \leq k \leq p-2$ and $0 \leq r \leq 2^{k}-1$ (see \cite{hl}). While this bound is tight for $p-k$ is small for $k<p/2$ it is not as good as the bound from \cite{af}, in (1).

The best general upper bound on $B(n)$ for even $n$, namely,
$$B(n)\leq\frac{n \lfloor \log n \rfloor}{2}
\eqno(3)
$$
obtained from the modified Kn\"{o}del graph (see \cite{bf},\cite{gf}). This bound, is better than the one in (1) for all even $n\neq 2^{p}$.

        In \cite{hx}, Harutyunyan and Xu presented an upper bound on $B(n)$ for odd $n$.
         They proved that for integers $n,p$, where $n>65$ is odd, $p \geq 7$ and $n\neq 2^{p}+1$,
$B(n) \leq \frac{(n+1)\lfloor {\log {n}} \rfloor}{2} + 2 \lceil \frac {n-1}{10} \rceil  - \lfloor {\frac{\lfloor {\log {n} } \rfloor +2}{4}}  \rfloor $.

However, recently Harutyunyan and Liestman presented in \cite{hln} a new upper bound for odd, positive integers, namely,

\begin{theorem}
\label{t1}
Let $n$ be an even integer such that $\lceil log n \rceil>2$ is prime, $m=\lceil log n \rceil \neq 2^{j}-1$ for any integer $j$, $m$ divides $n$, and for any integer $d \neq m-1$ which is a divisor of $m-1$, $2^{d} \not\equiv 1(mod (m))$. Then, $$B(n+1) \le \frac{n \lfloor log n \rfloor}{2} + \frac{n}{\lceil log n \rceil} + \lceil log n \rceil-2.
\eqno(4)
$$
\end{theorem}

         In this paper we present a new upper bound for $B(n)$, improving  the bounds in (1),(2),(3) and (4).
         Our main result is,
           \begin{theorem}
         \label{t2}
          Let $t,k,n$ be positive integers. Then, for a given $t \geq 7$ and $2 \le k \le \lfloor t/2 \rfloor -1$,
         \begin{enumerate}
          \item If $n=N=(2^{k}-1)2^{t+1-k}$,

          $$B(n) \le (k+1)N -(t-\frac{k}{2}+2)2^{k}+t-k+2.  \eqno(5.a)$$
          \item If $2^{t} < n<(2^{k}-1)2^{t+1-k}$,

          $$B(n) \le (k+1-p)n -(t-\frac{k}{2}+p+2)2^{k}+t-k -(p-2)2^{p}, \eqno(5.b)$$

          where $d=N-n$, $x= \lfloor \frac{d}{2^{t+1-k}} \rfloor$ and $ p = \left\{ \begin{array}{ll}
        \lfloor \log_{2}{(x+1)} \rfloor    & \mbox{if $x>0$ }\\
       0 & \mbox{otherwise.}\end{array}\right.
$

          \end{enumerate}


\end{theorem}

\section{Proof of Theorem \ref{t2}}

In this section we prove theorem \ref{t2}. First we construct a minimal broadcast graph and then demonstrate the broadcast scheme.

\subsection { Construction of the minimal broadcast graph}
We start by defining the binomial tree.
\begin{definition}

A binomial tree of order $t$, denoted by $B^{t}$, is defined recursively as follows:

A binomial tree of order $0$ is the trivial tree (a single vertex).

A binomial tree of order $t$ has vertex which is a root vertex whose children are roots of binomial trees of orders $t-1, t-2, ..., 2, 1, 0$ (in this order).

\end{definition}
\textbf{Observation:} The Binomial tree $B^{t}$ has $2^{t}$ vertices and height $t$.
Because of its unique structure, a binomial tree of order $t$ can be constructed
trivially from two trees of order $t-1$ by attaching one of them as the rightmost child of the root of the other one [see Figure.1].

\begin{figure}[H]
\centering
\includegraphics[width=3.5in]{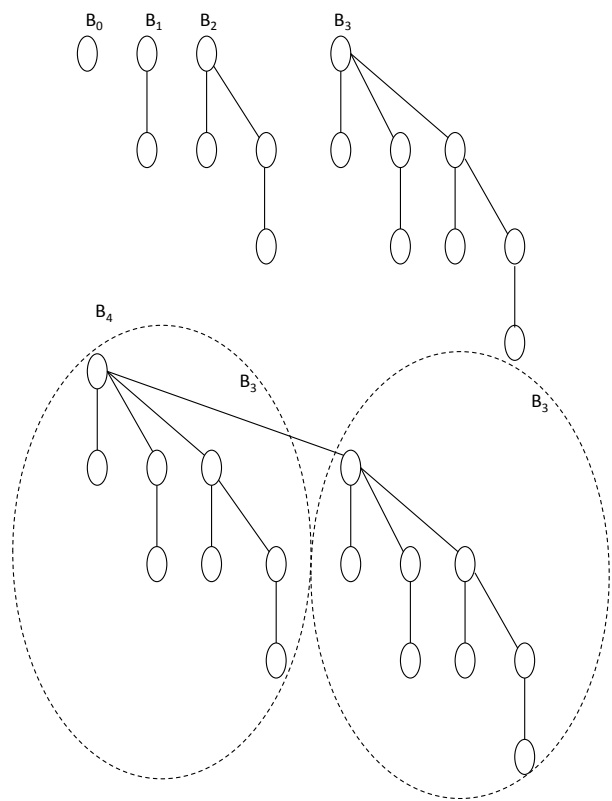}
\caption{The Binomial trees $B^{0}, B^{1}, B^{2}, B^{3}$ and $B^{4}$.}
\end{figure}

\begin{lemma}
\label{lemma1}
Let $B^{n}$ be the binomial tree of order $n$. Let $u$ be the root of $B^{n}$. Then, $b(u)=n$.
\end{lemma}

The proof is straightforward and is omitted.

Now we define a hypercube graph.

\begin{definition}
A hypercube graph of dimension $n$, denoted by $Q^{n}$, is defined recursively as follows:

A hypercube graph of dimension $0$ is a single vertex.

A hypercube graph of dimension $n$ is constructed of two hypercubes, each of dimension $n-1$, $Q^{n-1}_{1}$ and $Q^{n-1}_{2}$ and there is a perfect matching connecting the vertices of $Q^{n-1}_{1}$ with these of $Q^{n-1}_{2}$.

\end{definition}

\textbf{Notice:} A hypercube graph is a $n$-regular graph with $2^{n}$ vertices and thus has $n2^{n-1}$ edges.

\textbf{Observation:} Because of its unique structure, a hypercube graph of dimension $n$ can be constructed
trivially from  $n$ hypercube graphs of orders $n-1, n-2, ..., 2, 1, 0, 0$, denoted by $Q^{n-1}, Q^{n-2},.....,Q^{0},Q^{01}$, respectively.

$Q^{0}$ and $Q^{01}$ form a hypercube of dimension $1$,

$Q^{0},Q^{01}$ and $Q^{1}$ form a hypercube of dimension $2$,

etc...
\begin{figure}[H]
\centering
\includegraphics[width=3in]{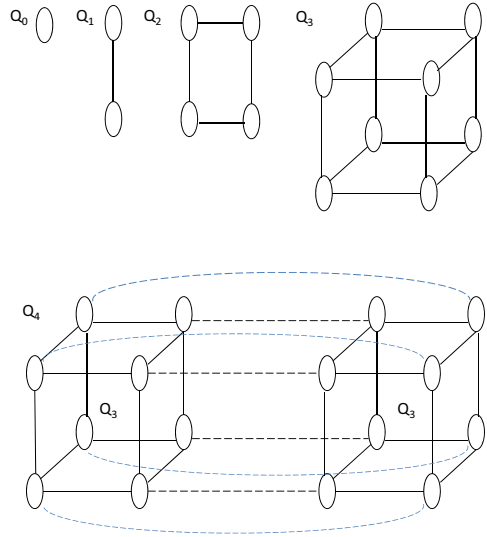}
\caption{The hypercubes $Q^{0}, Q^{1}, Q^{2}, Q^{3}$ and $Q^{4}$.}
\end{figure}

The following lemma is of great importance to our proof.
\begin{lemma}
\label{lemma2}
Let $Q^{n}$ be the $n$-dimensional hypercube. Then, for each vertex $u \in V(Q^{n})$, $b(u)=n$.
\end{lemma}

The proof is easy and follows by induction on $n$ and is omitted.

\textbf{Proof of theorem 1.2:}

First we demonstrate the construction of a broadcasting graph $G$ giving the upper bound of
$B(N)$ declared in (5.a), for $N=(2^{k}-1)2^{t+1-k}$ (case 1). The broadcasting graph  $G=(V,E), |V|=n$, with $2^{t} < n <N$ shall be constructed later (case 2). The broadcasting scheme in that graphs shall
demonstrate in the next section.

\textbf{Case 1:} For a given integer $t \ge 7$ and $k$, $2\le k\le \lfloor{t/2}\rfloor-1$, we construct a minimal broadcast graph $G=(V,E)$ with $|V|=N=(2^{k}-1)2^{t+1-k}$.


The broadcast graph $G$ is constructed of $2^{k}-1$ binomial trees
denoted  $B_{i}, 1 \le\ i \le 2^{k}-1$. Each $B_{i}$, $1 \le\ i \le 2^{k}-1$, is a $B^{t+1-k}$ tree.
Let $R=\{r_{1},...,r_{2^{k}-1}\}$ be the set of the roots of the binomial trees $B_1 , B_2, \ldots , B_{2^{k}-1} $,
respectively. For each $i$, $1 \le\ i \le 2^{k}-1$, $r_{i}$ is of degree $t+1-k$.

Thus, $ | V(G) | = N =(2^{k}-1)2^{t+1-k}$.

It is easily observed that $\lceil \log N \rceil=t+1$.

Denote by  $V_{1}=\{r_{1},...,r_{k-1}\}$ the set of the roots of the trees $B_{1},...,B_{k-1}$, respectively
and by $V_{2}=\{r_{k+1},...,r_{2^{k}-1}\}$ the set of the roots of the trees $B_{k+1},...,B_{2^{k}-1}$, respectively. Thus, $V_{1} \cup \{r_{k}\} \cup V_{2}=R$
 with $|R| = 2^{k} -1 $.

 We are ready now to construct the set $E(G)$ and to calculate its cardinality.
 First, we have the edges of the binomial trees.  Let $w \in B_{1}$ be the farthest leaf from the root $r_{1}$. We connect the vertices of $R \cup \{w\}$ in a way that they form a hypercube of dimension $k$, denoted by $Q^{k}$. Let $Q^{k-1}, Q^{k-2},.....,Q^{0},Q^{01}$, be the hypercube graphs that form $Q^{k}$ such that
 $w \in Q^{01}$ (in fact, $w=Q^{01}$) and for each $0 \le i \le k-1$, $r_{i+1} \in Q^{i}$. Let $Q^{k-1}_{1}$ and $Q^{k-1}_{2}$ be the two hypercube graphs of dimension $k-1$ that form $Q^{k}$ such that $Q_{1}^{k-1} = Q^{k-1}$ and $Q^{k-2},.....,Q^{0},Q^{01}$ form $Q^{k-1}_{2}$.
Now, we connect each vertex $v$, $v \in V\setminus (R\cup \{w\})$, in which its root, $r$, $r \in Q_{1}^{k-1}$, to each of the vertices in $V_{1} \cup \{r\}$.
For the vertices $v$, $v \in V\setminus (R \cup \{w\})$, in which their root $r$, $r \in Q_{2}^{k-1}$, we do the following: if $r \in Q^{i}$, $0 \le i \le k-2$, we connect $v$ to its root $r$, to each vertex in $V_{1}\setminus \{r_{i+1}\}$ and to $r_{k}$.

\textbf{Summary}: The $mbg$ graph $G$ constructed is a hypercube $Q^{k}$ of dimension $k$, and $2^{k}$ vertices (the set $R\cup \{w\}$), where each of the vertices in $R$ is a root of a binomial tree on $2^{t+1-k}$ vertices. Furthermore, each of the vertices of the binomial trees which are not on $R \cup \{w\}$ is adjacent to its root and to each of the vertices in $V_{1} \cup \{r_{k}\}$, except to $r_{j}$, if that vertex belongs to $Q^{j-1}$, for $1 \le j \le k$.

Now, we are ready to calculate the cardinality of $|E(G)|$.

First, the number of edges in the binomial trees is
 $$
 | \cup _{i=1}^{2^{k}-1} E(B_i )| = \sum _{i=1}^{2^{k}-1} |E(B_i )|=  (2^{k}-1) (2^{t+1-k} -1).
 \eqno(6)
 $$

 The number of edges in the hypercube induced on $R \cup \{w\}$ is
 $$
 |E(Q^{k})|=k2^{k-1}.
 \eqno(7)$$

The number of edges that connect each non root vertex in $G$ to its root is
$$
(2^{k}-1)[2^{t+1-k}-1-(t+1-k)]-1.
 \eqno(8)$$

The number of edges that connect the non root vertices in $Q_{2}^{k-1}\setminus\{w\}$ to $r_{k}$ is
$$(2^{k-1}-1)(2^{t+1-k}-1)-1.
 \eqno(9)$$

The number of edges that connect each vertex of $V_{1}$ to all vertices of $Q_{1}^{k-1}$ which are not roots (do not belong to $R$) is
$$(k-1)2^{k-1}(2^{t+1-k}-1).
 \eqno(10)$$

And finally, the number of edges that connect the vertices of $V_{1}$ to all the vertices in $Q_{2}^{k-1}\setminus \{w\}$ is
$$(k-2)(2^{k-1}-1)[(2^{t+1-k}-1)-1].
 \eqno(11)$$


Thus, summing the values in (6) up to (11) and recalling that $N=(2^{k}-1)2^{t+1-k}$ we obtain
 $$|E(G)|= (k+1)N -(t+2-\frac{k}{2})2^{k}+t+2-k.\eqno(12)$$

\textbf{Case 2:} We construct now a $mbg$ $G'=(V',E')$, $|V'|=n$, where $2^{t}<n<(2^{k}-1)2^{t+1-k}$. We start by constructing a $mbg$, $G=(V,E)$, with $|V|=N=(2^{k}-1)2^{t+1-k}$ as described in Case 1. Then, we obtain $G'$ from $G$ by deleting vertices and edges from $G$, in a way described below.

Define
$d=N-n$, $x= \lfloor \frac{d}{2^{t+1-k}} \rfloor$, $y=d-x2^{t+1-k}$ and

$ p = \left\{ \begin{array}{ll}
        \lfloor \log_{2}{(x+1)} \rfloor    & \mbox{if $x>0$ }\\
       0 & \mbox{otherwise.}\end{array}\right.$

Note that $0 \le x < 2^{k-1}$, $0 \le y < 2^{t+1-k}$ and $1 \le p < k$.

In order to construct $G'$ we delete vertices from $G$ as needed according to the value of $d$.
Since $d=2^{t+1-k}x+y$, the deletion process is done as follows:

 \begin{enumerate}
 \item If $x=0$, $d=y$, we delete $y$ vertices from some binomial tree in a way that we start deleting from the leaves and each vertex is deleted after all its descendants in the binomial tree are already deleted.

 \item If $x>0$, $d=2^{t+1-k}x+y$, we delete $2^{p}-1$ complete binomial trees and additional $2^{t+1-k}[x-(2^{p}-1)]+y$ non root vertices and then add $2^{p}-1$ edges. This is done in the following way:

    \begin{enumerate}
            \item Delete all the vertices that are in the binomial trees in which their roots form $Q^{0},Q^{1},....,Q^{p-1}$. Here, we delete $2^{p}-1$ binomial trees, where $p$ of these trees are rooted by vertices from $V_{1}$. Note that the hypercubes $Q^{0},Q^{1},....,Q^{p-1}$ are deleted from $Q^{k-1}_{2}$.

            \item Delete $2^{t+1-k}(x-(2^{p}-1))+y$ non root vertices from the trees in which their roots are in $Q_{2}^{k-1}\setminus (  \cup_{i=0}^{p-1} Q^{i})$. Note that since $p=\lfloor \log_{2}{(x+1)} \rfloor$, the number of vertices that we delete here is less than $2^{t+1-k}\cdot 2^{p}$.

            \item  For each vertex $b \in Q_{1}^{k-1} \cap R$, in which we have deleted its neighbor in $Q^{k-1}_{2}$, we connect $b$ to some vertex that remained in $Q^{k-2}$. Those edges that we add here replace the edges that connected $b$ to some other root in $Q^{k-1}_{2}$ that we have deleted in (a). This addition of edges is crucial in order to keep each vertex in the hypercube $Q^{k-1}_{1}$ matched to another vertex in $Q^{k-1}_{2}$.

    \end{enumerate}
\end{enumerate}

After the deletion process is ended we obtain in $G'$ the following sets: $R'$ is the set of the binomial trees roots. Then, $R' = V'_{1} \cup \{r_{k}\}\cup V'_{2}$, $|R'|=2^{k}-1-(2^{p}-1)=2^{k}-2^{p}$, where $V'_{1}=\{r_{1} ...r_{k-1-p}\}$, $|V'_{1}|=k-1-p$, $V'_{2} = R'\setminus (V'_{1}\ \cup \{r_{k}\})$ and $|V'_{2}| = 2^{k}-1-k-(2^{p}-1-p)=2^{k}-2^{p}+p-k$.

Now we calculate the number of edges that are deleted from $G$ in order to obtain the graph $G'$.

First, we count the edges that are adjacent to each non-root vertex in the $2^{p}-1$ complete binomial trees that were deleted from $G$.
The degree of each vertex $v$ in $V\setminus R$ is $k+j+1$, where $j$ is the distance of $v$ to the farthest leaf in its subtree. Indeed, $j$ edges connect $v$ to its direct siblings, $k$ edges connect $v$ to vertices in $R$ and one edge connects $v$ to its direct ancestor.
Since we delete a vertex after all its siblings are already deleted, the number of edges deleted each time we delete a vertex in $V \setminus R$ is $k+1$.
Therefore, the number of such edges that are deleted is
$$(k+1)(2^{t+1-k}-1)(2^{p}-1). \eqno(13)$$

Since the degree of each vertex in $Q^{k}$ is $k$, the number of edges that we delete from $Q^{k}$ is $k(2^{p}-1)$. By adding the $2^{p}-1$ edges, we actually omit from $Q^{k}$, as described, $$(k-1)(2^{p}-1)\eqno(14)$$ edges.

Note that if $x \neq 0$, the tree $B_{1}$ rooted in $r_{1}$ ($r_{1}=Q^{0}$) is deleted from $G$. Since $w \in B_{1}$, $w$ is deleted from $G$. The calculation in (14) includes the $k$ edges that connect $w$ to $Q^{k}$.

Now, we count the number of edges that connected the $p$ roots that were deleted from $V_{1}$ to all the non root vertices that remained in $G'$. This number is
$$[n-(2^{k}-2^{p})]p. \eqno(15)$$

Finally, we count $k+1$ edges for each of the $2^{t+1-k}(x-(2^{p}-1))+y$ non-root vertices that we delete from $Q^{k-1}_{2}$, which is:
$$(k+1)[2^{t+1-k}(x-(2^{p}-1))+y]. \eqno(16)$$

Summing (13)-(16), the total number of edges that we delete from $G$ in order to construct $G'$ is

$$np+(k+1)d-p2^{k}+(p-2)2^{p}+2.\eqno(17)$$

Now, by subtracting (17) from (12), recalling that $d=N-n$, we obtain that the  number of edges in $G'$:
$$|E(G')|=(k+1-p)n -(t-\frac{k}{2}+p+2)2^{k}+t-k -(p-2)2^{p}.\eqno(18)$$

This complete the proof of the construction of $mbg$ graph for $2^{t} <n \le N$.

\textbf{Observation:} One can easily observe that if $n=N$ and thus, $x=p=0$, we obtain $E(G')=(k+1)n -(t-\frac{k}{2}+2)2^{k}+t-k+2$ as in $(12)$.

\textbf{Remark:} For odd $n$ we can have $k \le \lceil \frac{t}{2} \rceil -1$.

\subsection {Broadcasting Scheme}
Let $u$ be an originator. We demonstrate a broadcasting scheme in the constructed graphs of cases 1 and 2.\\

\textbf{Case 1 :} $|V|=n=(2^{k}-1)2^{t+1-k}$.\\

\textbf{Case 1.1 :} Let $u \in R\cup \{w\}$. \\
The broadcasting scheme in that case is as follows:
Since the vertices of $R\cup \{w\}$ form a hypercube of $2^{k}$ vertices, at most $k$ time units are needed to complete broadcasting in $R\cup \{w\}$ (see lemma \ref{lemma2}).

\textbf{Case 1.2 :} $u \in Q_{1}^{k-1} \setminus R$. \\
At time unit $t=1$, $u$ transmits to its root, which needs another $k-1$ time units to accomplish broadcasting to all members of $Q^{k-1}_{1}$. At time unit $i$, $2 \le i \le k$, $u$ transmits to $r_{k-i+1}$ that needs another $k-i$ time units to accomplish broadcasting in $Q^{k-i}$. Broadcasting in $Q^{k-i}$ completes after time unit $k$ and therefore broadcasting in $Q^{k-1}_{2}$ completes at time unit $k$ (see lemma \ref{lemma2}). Therefore, broadcasting in $Q^{k}$ completes within $k$ time units.

\textbf{Case 1.3 :} $u \in Q_{2}^{k-1} \setminus (R\cup \{w\})$. \\
At the first time unit $u$ transmits the message to $r_{k}$, which needs another $k-1$ time units to accomplish broadcasting to all members of $Q^{k-1}_{1}$. Suppose $u\in Q^{j}$, $0 \le j \le k-2$. Then, at time unit $i$, $2 \le i \le k, i \neq j$, $u$ transmits the message to $r_{k-i+1}$ that needs another $k-i$ time units to accomplish broadcasting in $Q^{k-i}$ and thus, broadcasting in $Q^{k-i}$ completes after time unit $k$. At time unit $j$, $u$ transmits the message to its root that needs another $j$ time units to accomplish broadcasting in $Q^{j}$. Therefore, broadcasting in $Q^{k-1}_{2}$ completes at time unit $k$ and broadcasting in $Q^{k}$ complete within $k$ time units (see lemma \ref{lemma2}).

Now, in all three cases, after the first $k$ time units, each root in $R$ needs at most additional $t+1-k$ time units to complete broadcasting in its binomial tree (see lemma \ref{lemma1}). Thus, broadcasting in $G$ completes within at most $k+t+1-k=t+1$ time units, which is $b(u) \le t+1, \forall u \in V(G)$. \\

\textbf{Case 2 :} $2^{t}<n<(2^{k}-1)2^{t+1-k}$.\\
In this section we recall the definitions of $d,x$ and $p$ defined in case 2 in the previous section:
$d=N-n$, $x= \lfloor d/{2^{t+1-k}} \rfloor$, $p= \lfloor \log_{2}({x} +1)\rfloor$, where $0 \le x < 2^{k-1}$ and $0 \le p < k-1$.\\

\textbf{Case 2.1 :} $u \in R'$.\\
At the first time unit $u$ transmits the message to the other half of $Q^{k}$. Meaning, if $u \in Q_{2}^{k-1}$ then $u$ transmits the message to its neighbor in $Q^{k-1}_{1}$, or, $u \in Q_{1}^{k-1}$,and it transmits the message to its neighbor in $Q^{k-1}_{2}$. That is possible, since each vertex in $Q^{k-1}_{1}$ is connected to one of the vertices in $Q^{k-1}_{2}$.
Thus, after the first time unit $k-1$ more time units are needed to accomplish broadcasting in $Q^{k-1}_{1}$ and $Q^{k-1}_{2}$. Therefore, broadcasting in $R'$ is completing within at most $k$ time units.\\

\textbf{Case 2.2 :} $u \in Q_{1}^{k-1} \setminus R'$. \\
At time unit $t=1$, $u$ transmits to its root, that needs another $k-1$ time units to accomplish broadcasting to all members of $Q^{k-1}_{1}$. At time unit $i$, $2 \le i \le k-p$, $u$ transmits to $r_{k-i+1}$ that needs another $k-i$ time units to accomplish broadcasting in $Q^{k-i}$. Broadcasting in $Q^{k-i}$ completes after time unit $k$ and therefore broadcasting in $Q^{k-1}_{2}$ completes at time unit $k$ (see lemma \ref{lemma2}). Thus, broadcasting in $Q^{k}$ completes within $k$ time units.

\textbf{Case 2.3 :} $u \in Q_{2}^{k-1} \setminus (R' \cup \{w\})$. \\
At the first time unit $u$ transmits the message to $r_{k}$, which needs another $k-1$ time units to accomplish broadcasting to all members of $Q^{k-1}_{1}$. Furthermore, $u\in Q^{j}$, $p \le j \le k-2$. Then, at time unit $i$, $2 \le i \le k-p, i \neq j$, $u$ transmits the message to $r_{k-i+1}$ that needs another $k-i$ time units to accomplish broadcasting in $Q^{k-i}$. Thus, broadcasting in $Q^{k-i}$ completes after time unit $k$. At time unit $j$, $u$ transmits the message to its root that needs another $j$ time units to accomplish broadcasting in $Q^{j}$. Therefore, broadcasting in $Q^{k-1}_{2}$ completes at time unit $k$ and broadcasting in $Q^{k}$ complete within $k$ time units (see lemma \ref{lemma2}).

Now, in all three cases, each root in $R'$ needs at most $t+1-k$  additional time units to complete broadcasting in its binomial tree (see lemma \ref{lemma1}). Thus, broadcasting in $G'$ completes within at most $k+t+1-k=t+1$ time units

Hence, $b(u) \leq t+1, \forall u \in V(G')$.

This completes the proof of theorem \ref{t2}, in both cases.

\subsubsection{Example: Minimal broadcast network construction}
\begin{figure}[H]
\centering
\includegraphics[width=3.5in]{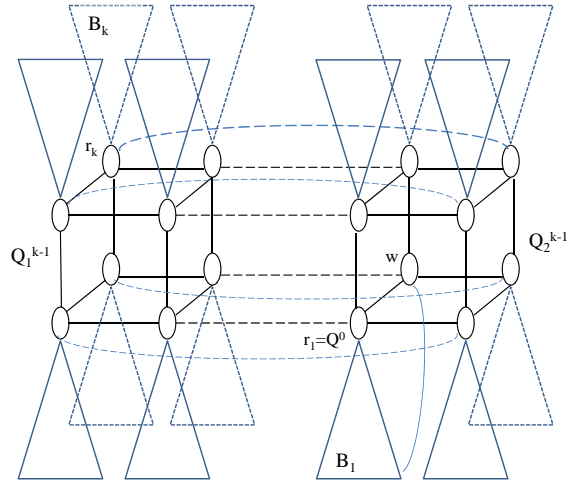}
\caption{This figure demonstrates the $mbg$ construction for $k=4$. The graph is constructed of $2^{k}-1$ binomial trees of dimension $t+1-k$. The set of binomial trees roots is $R$. The vertex $w$ is a leaf in $B_{1}$. The vertices in $R \cup \{w\}$ form a hypercube $Q^{k}$ of dimension $k$. The two hypercubes of dimension $k-1$ that form $Q^{k}$ are $Q_{1}^{k-1}$ and $Q_{2}^{k-1}$. Each vertex $v$, $v \in V\setminus (R\cup \{w\})$, in which its root, $r$, $r \in Q_{1}^{k-1}$, is connected to $k-1$ roots in $Q_{2}^{k-1}$ (the set $V_{1}$) and to $\{r\}$.
  Each vertex in $V\setminus (R \cup \{w\})$, in which its root $r$, $r \in Q_{2}^{k-1}$ and $r \in Q^{i}$, $0 \le i \le k-2$, is connected to its root $r$, to each vertex in $V_{1}\setminus \{r_{i+1}\}$ and to $r_{k}$.}
\label{fig03}
\end{figure}

\begin{table}[H]
\centering
\begin{tabular}{|c|c|c|c|c|c|c|c|c|c|}
 \hline
 $t$ & $k$ & maximal n & our result & \cite{hl} result & $t$ & $k$ & maximal n & our result & \cite{hl} result\\
 \hline

 7 &   2   &        192    &  551   &  557    &  15 &   2   &  49152    &             147407   &        147421    \\

 \hline

   8 &   2   &       384    &   1124   & 1131    &  15 &   3   &  57344    &             229266   &        229307    \\
 \hline
   8 &   3   &      448    &   1731   &  1751    &   15 &   4   &      61440    &     306973   &        307094    \\
 \hline

   9 &   2   &           768    &               2273   &          2281    &  15 &   5   &         63488    &             380476   &        380778
\\ \hline
   9 &   3   &           896    &               3516   &          3539    &  15 &   6   &         64512    &             450699   &        451375
\\ \hline

  10 &   2   &          1536    &               4574   &          4583    &  16 &   2   &         98304    &             294860   &        294875
\\ \hline
  10 &   3   &          1792    &               7093   &          7119    &  16 &   3   &        114688    &             458635   &        458679
\\ \hline
  10 &   4   &          1920    &               9448   &          9524    &  16 &   4   &        122880    &             614158   &        614288
\\ \hline

  11 &   2   &          3072    &               9179   &          9189    &  16 &   5   &        126976    &             761373   &        761698
\\ \hline
  11 &   3   &          3584    &              14254   &         14283    & 16 &   6   &        129024    &             902220   &        902949
\\ \hline
  11 &   4   &          3840    &              19033   &         19118    & 16 &   7   &        130048    &            1038539   &       1040073
\\ \hline

  12 &   2   &          6144    &              18392   &         18403    &  17 &   2   &        196608    &             589769   &        589785
\\ \hline
  12 &   3   &          7168    &              28583   &         28615    &  17 &   3   &        229376    &             917380   &        917427
\\ \hline
  12 &   4   &          7680    &              38218   &         38312    &  17 &   4   &        245760    &            1228543   &       1228682
\\ \hline
  12 &   5   &          7936    &              47257   &         47490    &  17 &   5   &        253952    &            1523198   &       1523546
\\ \hline

  13 &   2   &         12288    &              36821   &         36833    &  17 &   6   &        258048    &            1805325   &       1806107
\\ \hline
  13 &   3   &         14336    &              57248   &         57283    &  17 &   7   &        260096    &            2078796   &       2080445
\\ \hline
  13 &   4   &         15360    &              76603   &         76706    &  18 &   2   &        393216    &            1179590   &       1179607
\\ \hline
  13 &   5   &         15872    &              94842   &         95098    &  18 &   3   &        458752    &            1834877   &       1834927
\\ \hline

  14 &   2   &         24576    &              73682   &         73695    &  18 &   4   &        491520    &            2457328   &       2457476
\\ \hline
  14 &   3   &         28672    &             114585   &        114623    &  18 &   5   &        507904    &            3046879   &       3047250
\\ \hline
  14 &   4   &         30720    &             153388   &        153500    &  18 &   6   &        516096    &            3611598   &       3612433
\\ \hline
  14 &   5   &         31744    &             190043   &        190322    &  18 &   7   &        520192    &            4159437   &       4161201
\\ \hline
  14 &   6   &         32256    &             224970   &        225593    &  18 &   8   &        522240    &            4696076   &       4699666
\\ \hline

\end{tabular}
\vspace{3mm} \caption{In this table we show the number of edges for maximal values of $n=N=(2^{k}-1)2^{t+1-k}$ for $7 \le t \le 18$ and $2 \le  k \le \lfloor t/2 \rfloor -1 $. We compare our results with the results of \cite{hl}.}
\label{table1}
\end{table}

\begin{table}[H]
\centering
\begin{tabular}{|c|c|c|c|c|c|c|c|}

\hline
$n/|E(G')|$ & $ k=2$ & $ k=3$ & $ k=4$ & $ k=5$ & $ k=6$ &  \cite{hln} result & \cite{hl} result\\
\hline
16385  &   49109  &   49044  &   48909  &   48628  &   48043    &   115871 & \\
\hline
16386  &   49112  &   49047  &   48912  &   48631  &   48046    &    &    \\
\hline
16387  &   49115  &   49050  &   48915  &   48634  &   48049    &   115808   & \\
\hline
... & ... & ... & ... & ... & ... & ...&...\\
\hline
24575  &   73679  &   73614  &   73479  &   73198  &   72613    &   173670   &  \\
\hline
24576  &   73682  &   73617  &   73482  &   73201  &   72616    &     & \\
\hline
24577  &       &   98205  &   98080  &   97821  &   97284    &   173684  &    \\
\hline
24578  &     &   98209  &   98084  &   97825  &   97288    &       &    \\
\hline
24579  &       &   98213  &   98088  &   97829  &   97292    &   173698  & \\
\hline
...  & ... & ... & ... & ... & ... &...&...\\
\hline
28671  &        &  114581  &  114456  &  114197  &  113660    &   202615  &  \\
\hline
28672  &        &  114585  &  114460  &  114201  &  113664    &       &\\
\hline
28673  &        &        &  143153  &  142912  &  142413    &   202629  &  \\
\hline
28674  &        &        &  143158  &  142917  &  142418    &      & \\
\hline
... & ... & ... & ... & ... & ... &...&...\\
\hline
30719  &        &        &  153383  &  153142  &  152643    &   217087   &\\
\hline
30720  &      &  &  153388  &  153147  &  152648    &         &\\
\hline
30721  &      & &  &  183905  &  183440    &   217101  &\\
\hline
30722  &      & &   &  183911  &  183446    &        &\\
\hline
30723  &      & &  &  183917  &  183452    &   217116 & \\
 \hline
... & ... & ... & ... & ... & ... &...&...\\
 \hline
31743  &        &        &       &  190037  &  189572    &   224324   & \\
 \hline
31744  &      & &  &  190043  &  189578    &        &  \\
 \hline
31745  &      & & &  &  221393    &   224338  &  222016\\
 \hline
31746  &      & & &  &  221400    &        &  222023\\
 \hline
31747  &      & & &  &  221407    &   224352   & 222030\\
 \hline
 ... & ... & ... & ... & ... & ... &...&...\\
  \hline
32255  &        &        &        &        &  224963    &   227942  &  225586\\
 \hline
\end{tabular}
\vspace{3mm} \caption{In this table we show our result of $|E(G')|$ for $t=14$, $2 \le k \le 6$  and $ 16385 \le n \le 32255$. We compare our results with the results of \cite{hln} and \cite{hl}.}
\label{table4}
\end{table}






\begin{thebibliography}{20}

\bibitem{bh}
J. C. Bermond, P. Hell, A.L. Liestman, J. G. Peters. "Sparse broadcast graphs".\emph{ Discrete Applied Math.} 36(1992), 97-130.

\bibitem{bf}
J. C. Bermond, P. Fraigniaud, J. Peters. "Antepenultimate broadcasting". \emph{Networks,} 26(1995), 125-137.

\bibitem {dn}
M.J. Dinneen, M. R. Fellows, V. Faber. "Algebraic constructions of efficient broadcast networks". \emph{Applied Algebra, Algebraic Algorithms and Error Correcting Codes 9, Lecture Notes in Computer Science, vol} 539, Spring  Berlin(1991), 152-158.

\bibitem{dv}
M. J. Dinneen, J. A. Ventura, M. C. Wilson, G. Zakeri. "Compound construction of broadcast networks". \emph{Discrete Math.} 93(1999), 205-232.

\bibitem{af}
A. M. Farley. "Minimum broadcast networks". \emph{Networks} 9(1979), 313-332.


\bibitem{fh}
A. Farley, S. Hedetniemi, S. Mitchell, A. Proskurowski.
Minimum broadcast graphs.
{\it Discrete Math.} 25(1979), 189-193.

\bibitem{gf}
G. Fertin and A. Raspaud. "A survey on Kn\"{o}del Graphs". \emph{Discrete Applied Math.} 137(2004), 173-195.

\bibitem{gp}
M. Grigni and D. Peleg. "Tight bounds on minimum broadcast networks". \emph{SIAM J. on Discrete Math.} 4(1991), 207-222.

\bibitem{hl}
H. A. Harutyunyan, A. L. Leistman. "More broadcast graphs". \emph{Discrete Applied Math.} 98(1999), 81-102.

\bibitem{hh}
H. A. Harutyunyan: Minimum multiple message broadcast graphs. \emph{Networks} 47(4)(2006), 218-224.

\bibitem {hx}
Hovhannes Harutyunyan, Xiangyang Xu. New Construction of Broardcast Graphs. {\em 11th International Conference Information Visualization} (IV '07)(2007), 751-756.

\bibitem{hh}
H. A. Harutyunyan. "An efficient vertex addition method in broadcast networks".  \emph{Internet Math.} vol. 5, No. 3(2008), 211-225.

\bibitem{hln}
H. A. Harutyunyan, A. L. Leistman. "Upper bounds on the broadcast function using minimum dominating sets".  \emph{Discrete Math.}  312(2012), 2992-2996.

\bibitem{lh}
L. H. Khachatrian and H. S. Haroutunian. "Constuction of new classes of minimal broadcast networks". in Proc. \emph{3rd International Colloquium on Coding Theory,} Armenia(1990), 69-77.

\bibitem {WK}
W. Kn\"{o}del. "New gossips and telephones". \emph{Discrete Math.} 13(1975), 95.

\bibitem {mh}
S. Mitchell and S. Hedetniemi.
"A census of minimum broadcast graphs". \emph{J. Combin. Inform. Systems Sci.} 5(1980), 141-151.


\end{thebibliography}
\end{document}